\documentclass{article}

\usepackage[dvips]{graphicx}

\begin{document}

\title{A simple method to determine
parameters of  embryos distribution in
homogeneous nucleation under dynamic conditions
}
\author{Victor Kurasov}
\date{Victor.Kurasov@pobox.spbu.ru}

\maketitle

The result of construction of nucleation kinetics
in dynamic conditions
\cite{Kinetics}
can be accumulated in the following: {\it
To construct kinetics
it is  necessary to solve the balance equation
\begin{equation}
\Phi = \zeta + g
\ \ , \ \ \ \ \
g(z) =
\int_{-\infty}^{z} dx (z-x)^{3} f(x) \ \ .
\end{equation}
The value of the ideal supersaturation $\Phi$
 can be presented in the form
\begin{equation}\label{4.2.h2}
\Phi = \Phi_{*} + \Phi_{*} ( cx + lx^2)/\Gamma
\end{equation}
with two parameters
\begin{equation}
c=\frac{\Gamma}{\Phi_{*}} \frac{d\Phi}{dx}\mid_{x=0} \ \ , \ \ \
l=\frac{\Gamma}{2 \Phi_{*}} \frac{d^2\Phi}{dx^2}\mid_{x=0}   \leq 0   \ \ .
\end{equation}
The approximation
\begin{equation}
f(x) = f_{s} \exp(\Gamma \frac{\zeta(x) - \Phi_{*}}{\Phi_{*}})
\ \ , \ \ \ \ \ f_s = f_s(\zeta) \mid_{\zeta = \Phi_*}
\end{equation}
is valid during nucleation period.
Index $*$ corresponds to the maximum of
supersaturation.}

Analyze the behavior of $g(z)$.
Extract three regions.
Let  $l \sim
0 $.
Extract the period of the  appearing of the main consumers of the vapor
during the formation of the main quantity of  droplets. The behavior
of
$g$
in the first approximation can be seen from
\begin{equation}
g(z) \sim \exp(z) \int_{0}^{\infty} y^{3} \exp(-y)dy \ \ .
\end{equation}
One can see that
{\it
The subintegral function in the last expression is essential only when
\begin{equation}\label{4.2.h4'}
2 \sim  y_{min} \leq  y \leq y_{max} \sim 4 \ \ ,\ \ \ \
y= c\rho \ \ .
\end{equation}
This period corresponds to the region of the appearing
of the main consumers of the vapor in the current moment.
When the last equations aren't valid the subintegral function is negligible.
The period corresponding to  (\ref{4.2.h4'})
plays the main role in the formation of the spectrum.
}
In nucleation one can extract  the initial region according to
\begin{equation}
-5 \leq cz \leq c z_{b}  \ \ .
\end{equation}
The value of  $z_b$
has to be determined from
\begin{equation}
\frac{f(x) - f_{1}(x) }{ f_{1}(x)} \ll 1 \ \ ,
\end{equation}
where $f$
is the distribution of the droplets  and
$f_{1}$
is the imaginary distribution obtained on the base of
$\Phi$
instead of  $\zeta$.
The values in this region will be marked by the index "$in$".
 The period of nucleation is finished
when
$cz \sim 1$.
It starts when $cz \sim -1$.
One can define the peak of nucleation according to
\begin{equation}
- c^{-1} < z < c^{-1} \ \ .
\end{equation}
 Also one can conclude that
$$
- y_{min} \leq cz_{b}
\ \ ,
- y_{min} + 1 \leq c z_b
\ \ .
$$

{\bf
Approximation of the spectrum
}

It is necessary to construct some approximation for $f(x)$.
Then it is necessary to get  $\zeta - \zeta_{*}$
in the standard approximation
\begin{equation}
f(x) = f_{*} \exp(\frac{\Gamma (\zeta - \zeta_{*})}{\Phi_{*}})
\exp(\frac{\Gamma (\zeta_{*} - \Phi_{*})}{\Phi_{*}})
\ \ .
\end{equation}

One can decompose $\zeta - \zeta_{*}$
 in the neighborhood of
$z=0$.
The expressions for the derivatives will be the following
\begin{equation}
\frac{d\zeta}{dz} = \frac{d\Phi}{dz} - 3 \int_{-\infty}^{z}
f(x) (z-x)^2 dx \ \ ,
\end{equation}
\begin{equation}
\frac{d^2\zeta}{dz^2} = \frac{d^2\Phi}{dz^2} - 6 \int_{-\infty}^{z}
f(x) (z-x) dx        \ \ ,
\end{equation}
\begin{equation}
\frac{d^3\zeta}{dz^3} = \frac{d^3\Phi}{dz^3} - 6 \int_{-\infty}^{z}
f(x)  dx                  \ \ ,
\end{equation}
\begin{equation}
\frac{d^n\zeta}{dz^n} = \frac{d^n\Phi}{dz^n} - 6 \frac{d^{n-4} f(z)}{dz^{n-4}}
\ \ , \ \ \ \ \ \ \ n \geq 4 \ \ .
\end{equation}

The value  $d^3 g / d z^3 $
is proportional to the total number of the droplets.
 According to the iteration
procedure one can note that
{\it
At the first step of the iterations \cite{Iterations} the
relative error  goes to
$\infty$.
On the second step the characteristics are determined with a high accuracy.
}
So, the account of the last two  expression  doesn't  lead  to  any
essential
influence. Then one can restrict  by the two first terms
\begin{equation}\label{4.2.h12}
f(x) = f_{*} \exp(\Gamma \frac{ d^2 \zeta}{2dz^2}\mid_{z=0}
\frac{x^2} {\Phi_{*}})
 \exp(\frac{\Gamma (\zeta_* - \Phi_{*})}{\Phi_{*}})
\ \ .
\end{equation}
This approximation is  valid only near the peak of nucleation. The standard
method of the steepens descent spreads this approximation on the whole
period of nucleation (including the initial region). This leads to the
essential relative error.
At the same time any approximation at the initial region isn't necessary
because the  write approximation of the spectrum here is already established
\begin{equation}\label{4.2.h13}
f(x) = f_{*} \exp(\Gamma \frac{\Phi(x) - \Phi_{*}}{\Phi_{*}})
\ \ .
\end{equation}

{\bf
The equations on the parameters of the spectrum
}

To get the equation on the parameters of the spectrum one can differentiate
the balance equations at $z=0$ and get
\begin{equation}
\Phi(0) = g_{in}(0) + g_{ex}(0) +\zeta(0) \ \ ,
\end{equation}
\begin{equation}
\frac{d\Phi}{dz}\mid_{z=0}  =\{ \frac{dg_{in}}{dz} +\frac{d g_{ex}}{dz} +
\frac{d\zeta}{dz} \} \mid_{z=0}                \ \ ,
\end{equation}
\begin{equation}
\frac{d^2\Phi}{dz^2}\mid_{z=0}  =\{ \frac{d^2g_{in}}{dz^2}
+\frac{d^2 g_{ex}}{dz^2} +
\frac{d^2\zeta}{dz^2}  \} \mid_{z=0}                \ \ ,
\end{equation}
The indexes  "$in$"
and "$ex$"
marks the parts of  $g$
corresponding to the droplets formed in the "initial" period and outside
it. The relations
(\ref{4.2.h13})
and (\ref{4.2.h2})
lead to
\begin{equation}
g_{in} = \sum_{i=0}^{3} z^{3-i} \alpha_{i} \ \ ,
\ \
\frac{d g_{in}}{dz} = \sum_{i=0}^{2} z^{2-i} (3-i) \alpha_{i}
\ \ ,
\end{equation}
\begin{equation}
\frac{d^2 g_{in}}{dz^2} = \sum_{i=0}^{1} z^{1-i} (3-i)(2-i) \alpha_{i}
\ \ .
\end{equation}
In these equations $\alpha_{i}$
are defined according to
\begin{equation}
\alpha_{i} = (-1)^{i} \frac{3!}{i!(3-i)!} f_{*} \int_{-\infty}^{z_{b}}
x^{i} \exp(cx+lx^2) dx
\ \ .
\end{equation}
To obtain the algebraic system  one can substitute the integral
term
in the last expression by one of the Boyd estimates
\begin{eqnarray}\label{4.2.Boyd}
\frac{\pi/2}{\sqrt{z^2+\pi}+(\pi-1) z} \leq \exp(z^2)
\int_{z}^{\infty} \exp(-t^2) dt \leq
\frac{\pi/2}{\sqrt{(\pi-2)z^2+\pi}+ 2 z}
\\ \nonumber
\ \ \ z>0
\end{eqnarray}
Analogous procedure can be applied for the values
$g_{ex}$, $\frac{dg_{ex}}{dz}$,
 $\frac{d^2g_{ex}}{dz^2}$.
According to
 (\ref{4.2.h12})
the expressions for
$$g_{ex}(0) \ \ \ dg_{ex}/dz\mid_{z=0} \ \ \
d^2 g_{ex}/dz^2 \mid_{z=0}$$
have the following form
\begin{equation}
g_{ex}(0) = f_{m} \int_{z_{b}}^{0} \exp(-(\frac{x}{x_{p}})^2) x^3 dx =
f_{m} x_{p}^4 \lambda_{3}
\ \ ,
\end{equation}
\begin{equation}
\frac{dg_{ex}}{dz} \mid_{z=0}= 3 f_{m}
\int_{z_{b}}^{0} \exp(-(\frac{x}{x_{p}})^2) x^2 dx =
3 f_{m} x_{p}^3 \lambda_{2}
\ \ ,
\end{equation}
\begin{equation}
\frac{d^2g_{ex}}{dz^2} \mid_{z=0}= 6 f_{m}
\int_{z_{b}}^{0} \exp(-(\frac{x}{x_{p}})^2) x dx =
6 f_{m} x_{p}^2 \lambda_{1}
\ \ ,
\end{equation}
where
\begin{equation}
\lambda_{i} = \int_{\xi_{0}}^0 \xi^i \exp(-\xi^2) d\xi \ \ , \ \ \
\xi_{0} = \frac{z_{b}}{x_{p}}
\end{equation}
and
$x_p$
is some parameter like the characteristic half-width.

As the result one can see the system of the three algebraic equations which
can
be solved by the ordinary numerical methods.

{\bf
Linear source
}

In the case
 $l=0$
the final expressions become more simple. For
$z_{b}$
one
can take here the value
$-x_{p}$.
Then
taking into account
$\Gamma \gg 1$,
 one can obtain
\begin{equation}\label{4.2.h111}
\alpha_{i} = f_{*} \frac{3!(-1)^i}{i!(3-i)!} c^{-i-1} \sum_{j=0}^{i}
\frac{i!}{(i-j)!} (-1)^{j} (-cx_{p})^{i-j} \exp(-cx_{p})
\ \ .
\end{equation}
This induces the following expressions
\begin{equation}
g_{in}(0) = \alpha_{3} = -  f_{*} c^{-4} \sum_{j=0}^{3}
\frac{3!}{(3-j)!} (-1)^{j} (-cx_{p})^{3-j}
\exp(-cx_p) \ \ ,
\end{equation}
\begin{equation}
\frac{dg_{in}}{dz}\mid_{z=0} = \alpha_{2}
 =  f_{*} 3 c^{-3} \sum_{j=0}^{2}
 (-1)^{j} \frac{2!}{(2-i)!} (-cx_{p})^{2-j}
\exp(-cx_p)      \ \ ,
\end{equation}
\begin{equation}\label{4.2.h112}
\frac{d^2g_{in}}{dz^2}\mid_{z=0} =2 \alpha_{1}
 = -  f_{*} 6 c^{-2} \sum_{j=0}^{1}
 (-1)^{j} (-cx_{p})^{1-j} \exp(-c x_p)
\ \ .
\end{equation}
The values of
 $\lambda_{i}$
become the universal constants
\begin{equation}
\lambda_{i} = \int_{-1}^{0} \exp(-x^2) x^i dx
\ \ .
\end{equation}

Introduction of the connection between
 $\frac{d^2\zeta}{dz^2}$
and
$x_{p}$
\begin{equation}
\mid \frac{d^2\zeta}{dz^2} \mid = \frac{2 \zeta_{*}}{\Gamma x_{p}^2}
\approx \frac{2 \Phi_{*}}{\Gamma x_{p}^2}
\end{equation}
leads to the transformation from
(\ref{4.2.h111}) - (\ref{4.2.h112})
to the following form
\begin{equation}\label{4.2.h115}
\Phi_{*} = \zeta_{*} - f_{*} c^{-4} \sum_{j=0}^{3} \frac{3!}{(3-j)!}
(-1)^{j} (-c x_{p})^{3-j} \exp(-cx_p) + f_{m} x_{p}^4 \lambda_{3}
\ \ ,
\end{equation}
\begin{equation}\label{4.2.h116}
\frac{\Phi_{*}c}{\Gamma}
 = f_{*} c^{-3} 3 \sum_{j=0}^{2} \frac{2!}{(2-j)!}
(-1)^{j} (-c x_{p})^{2-j}
\exp(-c x_p)  + 3 f_{m} x_{p}^3 \lambda_{2}
\ \ ,
\end{equation}
\begin{equation}\label{4.2.h117}
0 =   - f_{*} c^{-2} 6 \sum_{j=0}^{1}
(-1)^{j} (-c x_{p})^{1-j}
\exp(-c x_p)  + 6 f_{m} x_{p}^2 \lambda_{1} - \frac{2 \Phi_{*}}
{\Gamma x_{p}^2}
\ \ .
\end{equation}

Then
\begin{equation}\label{4.2.h118}
f_{*} =
\frac{\Phi_{*}c}{\Gamma}
[ c^{-3} 3 \sum_{j=0}^{2} \frac{2!}{(2-j)!}
(-1)^{j} (-c x_{p})^{2-j}
\exp(-c x_p)  + 3 \Psi x_{p}^3 \lambda_{2}]^{-1}
\ \ ,
\end{equation}
where
$$
\Psi = \frac{f_m}{f_*} \ \ ,
$$
and
(\ref{4.2.h117})
transforms to
\begin{equation}\label{4.2.h119}
-
\frac{ \Phi_* c \{ c^{-2} 6 \sum_{j=0}^{1}
(-1)^{j} (-c x_{p})^{1-j} \exp(-c x_p)  + 6  x_{p}^2 \lambda_{1} \Psi \}
}
{
\Gamma
\{ c^{-3} 3 \sum_{j=0}^{2} \frac{2!}{(2-j)!}
(-1)^{j} (-c x_{p})^{2-j}\exp(-c x_p)  + 3 \Psi x_{p}^3 \lambda_{2} \}
}
 = \frac{2 \Phi_{*}}
{\Gamma x_{p}^2}
\ \ .
\end{equation}

The system
(\ref{4.2.h115})-(\ref{4.2.h117})
allows some modifications which are
available for the nonlinear case also.
Note that the subintegral function
 $f(x)x$
in the expression for
$d^2 g /dz^2$
is
essential only in the  finite region of nucleation. The approximation
(\ref{4.2.h12})
also leads to the same conclusion.
Then the approximation
(\ref{4.2.h12})
can be formally spread to the whole
nucleation period. Then the  equation
(\ref{4.2.h117})
 leads to the following expression for
$x_{p}$
\begin{equation}
x_{p} = (\frac{2 \Phi_{*}}{3 \Gamma f_{m}} )^{1/4}
\ \ .
\end{equation}
In the  nonlinear case the  equation
(\ref{4.2.h117})
 has the following form
\begin{equation}
\frac{d^2 \Phi }{dz^2} = 3 x_{p}^2 f_{m} - 2 \frac{\Phi_{*}}{\Gamma x_{p}^2}
\ \ .
\end{equation}
This bisquare equation gives
the $x_{p}$
dependence on
 $f_{m}$.
Then the second
equation of the system
(\ref{4.2.h115}) - (\ref{4.2.h117})
becomes the closed
equation  and can be solved by iterations. These
iterations
can take into account  the sharp dependence of
$\exp(-F_{c})$
 on the  supersaturation.
The zero approximation can be chosen as
 $\Gamma = 1, \Phi_{*} = 1$.
Then the second
approximation gives a rather precise result for all parameters
of the  spectrum. Certainly, the amplitude
 $f_{m}$
must be obtained by
the next (hidden) iteration
instead of the explicit expression in the classical theory of nucleation.

The system of condensation equations can be even more simplified if one
notices that the approximation
(\ref{4.2.h13})
can be spread
over the whole nucleation
in the first and in the
second equations of the system.  Finally, the system
of condensation equations has the following form
\begin{equation}\label{4.2.h221}
\frac{d\Phi}{dz}\mid_{z=0} =  3 f_{*} \int_{-\infty}^{0}
x^2 \exp(cx+lx^2) dx
\ \ ,
\end{equation}
\begin{equation}\label{4.2.h222}
\frac{d^2 \Phi }{dz^2} = 6 \hat{\lambda}_{1}
 x_{p}^2 f_{m} - 2 \frac{\Phi_{*}}{\Gamma x_{p}^2}
\ \ ,
\ \ \ \ \
\hat{\lambda}_{1} = \int_{-\infty}^{0} y \exp(-y^2) dy = 0.5
\ \ .
\end{equation}

One must take into account the difference between
 $f_{m}$
and
 $f_{*}$,
which is given by
the first equation of the system where
\begin{eqnarray}
\frac{f_{*}}{f_{m}} =
 \exp(
-
\frac{\Gamma d\Phi}{3\Phi_* dz}\mid_{z=0}
\int_{-\infty}^{0}
x^3 \exp(cx+lx^2) dx / \int_{-\infty}^{0}
x^2 \exp(cx+lx^2) dx )
\ \ .
\end{eqnarray}

It is essential that the  equations
(\ref{4.2.h221})
and
 (\ref{4.2.h222})
are the  separate
ones. The equation
 (\ref{4.2.h221})
can be solved by iterations. The total number of
the droplets can be obtained as
\begin{equation}
N = f_{m} \sqrt{\pi} x_{p}
\ \ .
\end{equation}

One can prove that
the relative error of the last expression
$\delta$
increases when
$l$
decreases in the absolute value
\begin{equation}
\frac{d \mid \delta \mid }{dl} > 0
\ \ .
\end{equation}
In the linear case
\begin{equation}
g(0) = \frac{\Phi_{*}}{\Gamma}
\ \ ,
\ \
x_{p} = c^{-1} (4\exp(1))^{1/4}
\ \ ,
\end{equation}
\begin{equation}
N = \frac{\Phi_{*} c^3 (4\exp(1))^{1/4} \sqrt{\pi}}{6 \Gamma \exp(1)}
\ \ .
\end{equation}
Then
$N$
is
$1.19$
times greater than the result of the iteration procedure
at the second step and
$1.03$
times greater than the result
at the third step. It can be regarded as the practically precise result.
All the estimates obtained for
$l=0$
remain valid in the non
linear case.

\end{document}